\documentclass[prl,superscriptaddress,amsmath,amssymb,longbibliography,12pt]{revtex4-2}

\usepackage[utf8]{inputenc}
\usepackage[T1]{fontenc}
\usepackage{lmodern}
\usepackage{subcaption}
\usepackage{caption}
\usepackage{stackengine}
\usepackage{amsmath}
\usepackage{amssymb}
\usepackage{enumitem}
\usepackage{graphicx}
\usepackage{braket}
\usepackage{bm}
\usepackage{algorithm}
\usepackage{algpseudocode}
\usepackage{float}
\usepackage{xcolor}
\usepackage[colorlinks=true,citecolor=blue,urlcolor=blue, linkcolor=blue]{hyperref}

\begin{document}

\title{Efficient encoding of the 2D toric code logical state using local Clifford gates}

\author{Ivan H.C. Shum}
\email{hcs64@cam.ac.uk}
\affiliation{Department of Applied Mathematics and Theoretical Physics, University of Cambridge}

\begin{abstract}
An algorithm which encodes the $L\times L$ 2D toric code logical state with a circuit of depth $2L+1$, using only local controlled-NOT($CX$) and Hadamard($H$) gates, is presented. 
\end{abstract}

\maketitle

\tableofcontents

\section{Introduction}

It is well known that one cannot prepare the logical state (or any ground state) of the 2D toric code with a circuit of sublinear depth due to a Lieb-Robinson bound \cite{Bravyi2006}. Previous attempts at using local Clifford gates and no measurements to encode the logical state of the toric code have worked in the Schrodinger picture of the problem \cite{Higgott2021, Chen2024}; in this paper, a diagrammatic description of working in the Heisenberg picture is provided, similar to the framework described in \cite{Gibbs2025}, and a circuit that encodes the logical state using local Clifford gates; that is, the two sites a $CX$ gate acts on have Manhattan distance no greater than one; and is of lower depth than those presented in \cite{Higgott2021, Chen2024}. 

\section{Notation}

An arrow will be used to represent $CX$ gates, with the tail marking the control qubit and the arrowhead pointing at the target qubit. Qubits will be marked by black dots. \\

Lone X terms will be marked by a red cross, while lone Z terms will be marked by a blue circle. Due to the nature of the algorithm, which only uses $CX$ gates before the last step, X interactions are mapped to X interactions and Z interactions are mapped to Z interactions. Hence a chain of red/blue segments will be used to represent the X/Z interactions respectively. \\

Without loss of generality, when with no further specification $CX$ denotes the gate with control on first qubit and target on the second. Denoting the Pauli matrices by $\{I, X, Y, Z\}$ and noting $CX^\dagger=CX, CX^2=II$, one has the following conjugation relations: \\
$CX(IX)CX=IX, CX(XX)CX=XI, CX(XI)CX=XX$\\
$CX(ZI)CX=ZI, CX(ZZ)CX=IZ, CX(IZ)CX=ZZ$\\

\begin{figure}[H]
    \centering
    \includegraphics[width=3cm]{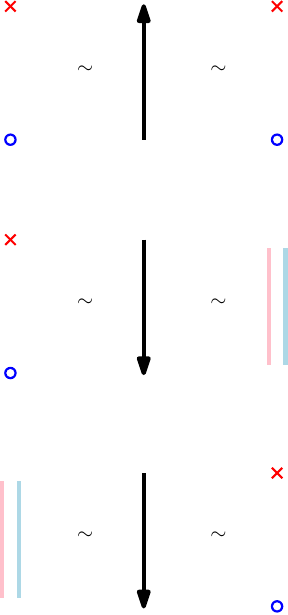}
    \caption{Basic operations after conjugation of $CX$ gates. }
\end{figure}

 The coordinate system used is a subset of the lattice points of $[0, 2L-1]\times [0,2L-1]$ for the $L\times L$ toric code, with arithmetic done mod $2L$, identifying sites with even parity ($(x,y)$ where $x+y$ is even) hosting qubits, sites with odd parity and the first coordinate even hosting a plaquette/$Z$ operator centred at it, which means given $(2i+1, 2j)$, there is a $Z$ operator with support on $(2j, 2j), (2j+2, 2j), (2j+1, 2j-1), (2j+1, 2j+1)$, and sites with odd parity and the second coordinate being even hosting a star/$X$ operator centred at it. 

 \begin{figure}[H]
    \centering
    \includegraphics[width=10cm]{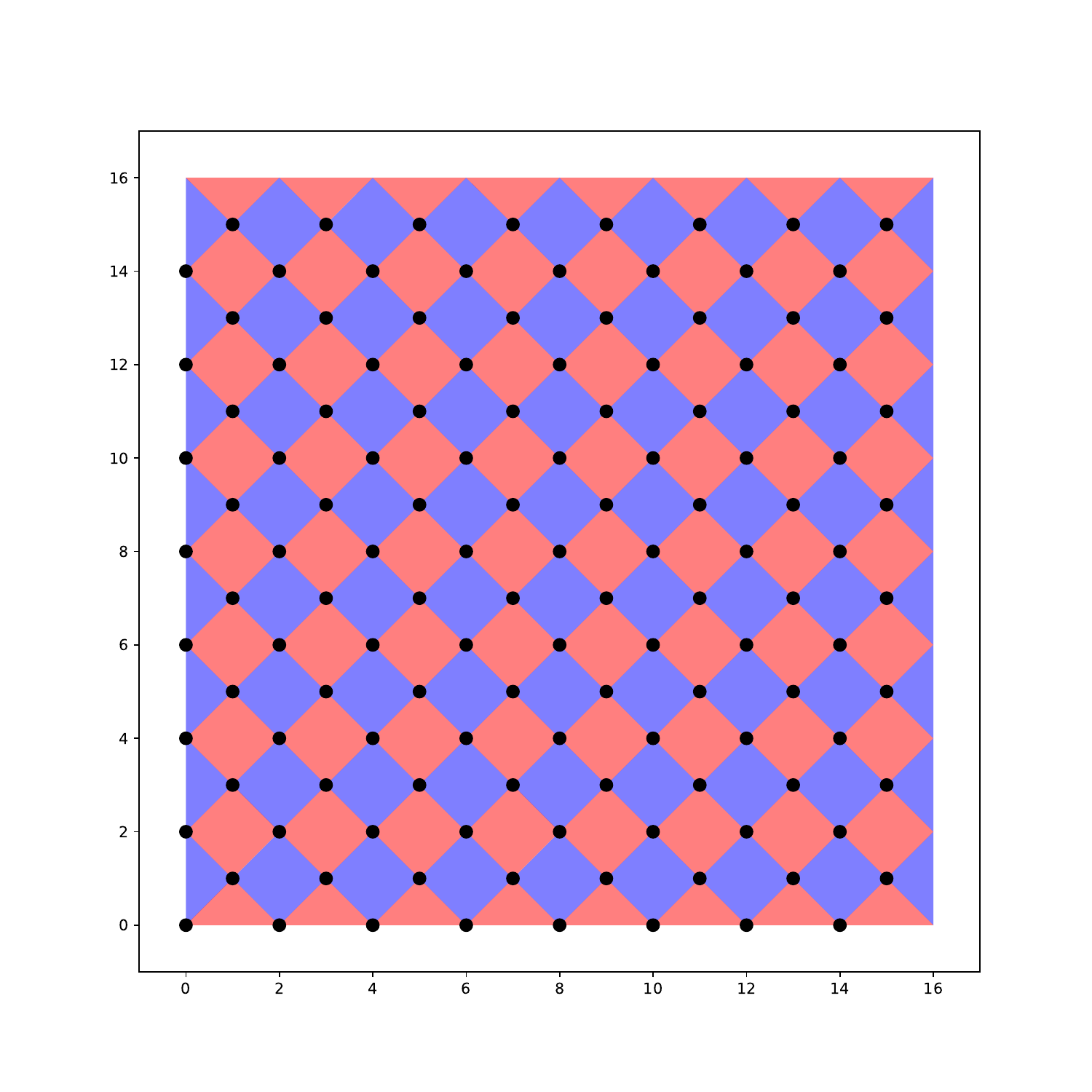}
    \caption{Coordinate system used to represent the 2D toric code. Periodic boundary conditions are implied. Qubits are shown without repetition while the triangles on the top/bottom and left/right should be interpreted as one 4-site operator. }
\end{figure}

 Defining $s(x_1, x_2)=\min(|x_1-x_2|/2, L-|x_1-x_2|/2)$, the distance between two qubits is then computed as $d((x_1, y_1), (x_2, y_2))=s(x_1, x_2)+s(y_1, y_2)$, such that each qubit has 8 neighbours. 

\section{Description of operations and previous results}

The idea presented in this paper is to unitarily transform the Hamiltonian $H=-\sum A_v-\sum B_p$ of the toric code, where $A_v$ and $B_p$ are the star/X and plaquette/Z operators respectively, to a simpler form $\tilde{H}=UHU^\dagger=-\sum C_i -\sum D_j$, where $C_i$ is an indexed set of $X$ operators, $D_j$ is an indexed set of $Z$ operators, and they act on disjoint sets of qubits non-trivially, called the $X$ type and $Z$ type qubits respectively, while $\tilde{H}$ also acts on 2 qubits trivially, called the trivial qubits. \\

The unitary $U$ will be chosen to only consist of layers of commuting $CX$ gates, due to the fact that they map purely $X$ interactions to purely $X$ interactions, and purely $Z$ interactions to purely $Z$ interactions. \\

Then the ground state is clearly fourfold degenerate, and is also partially separable: a basis for the ground eigenspace is \[\{\ket{00}, \ket{01}, \ket{10}, \ket{11}\}\bigotimes\ket{0}^{\otimes L^2-1}\bigotimes\ket{+}^{\otimes L^2-1}\] where the first term is defined on the two trivial qubits, the second term is defined over the $Z$ type qubits and the third term is defined over the $X$ type qubits. \\

In particular, the logical computational basis states of the toric code, say $\{\ket{00_L}, \ket{01_L}, \ket{10_L}, \ket{11_L}\}$, are mapped to the ground eigenspace of $\tilde{H}$ correspondingly, if one can show that a representative of $X_{L_1}, Z_{L_1}, X_{L_2}, Z_{L_2}$ are mapped to the four single site $X$ and $Z$ operators defined over the 2 trivial qubits. Inverting the process will prepare the logical state. \\

\begin{table}[H]
\begin{center}
\begin{tabular}{|c|c|c|c|}
\hline
Hamiltonian & $-\sum A_v-\sum B_p$ & $-\sum C_i-\sum D_j$ & $-\sum HC_iH-\sum D_j$ \\
\hline
State & $a\ket{00_L}+b\ket{01_L}+c\ket{10_L}+d\ket{11_L}$ & $\ket{\psi}\bigotimes\ket{0}^{\otimes L^2-1}\bigotimes\ket{+}^{\otimes L^2-1}$ & $\ket{\psi}\bigotimes\ket{0}^{\otimes 2(L^2-1)}$\\
\hline
\end{tabular}
\end{center}
\caption{The equivalent operations on Hamiltonian and state during an encoding process. Denote the 2-qubit state to be encoded as $\ket{\psi}=a\ket{00}+b\ket{01}+c\ket{10}+d\ket{11}$. Note the Hadamard gates only act on $X$ type qubits. Previous papers \cite{Higgott2021, Chen2024} evolve along the second row, right to left; this paper evolves along the first row, left to right. }
\end{table}

Note that the underlying idea is nothing new: it is simply a graphical representation of the idea presented in \cite{AaronsonGottesman2004}, which also provides a simpler, but much less efficient, algorithm for general stabilizer codes that only consist of purely $X$ and $Z$ operators. \\

Performing Gaussian elimination on a tableau for a stabilizer codes with $O(L^2)$ qubits, translates into a quantum algorithm with $O(L^6)$ depth and non-local gates in the most general case; the ideas presented here improve the depth to $2L+1$ and only use local gates. \\

\section{Main algorithm}

The code is attached at \href{https://github.com/ihcs64/toricCodePreparation}{https://github.com/ihcs64/toricCodePreparation}, using the Stim library \cite{Gidney2021}. 

\begin{figure}[H]
    \centering
    \includegraphics[width=10cm]{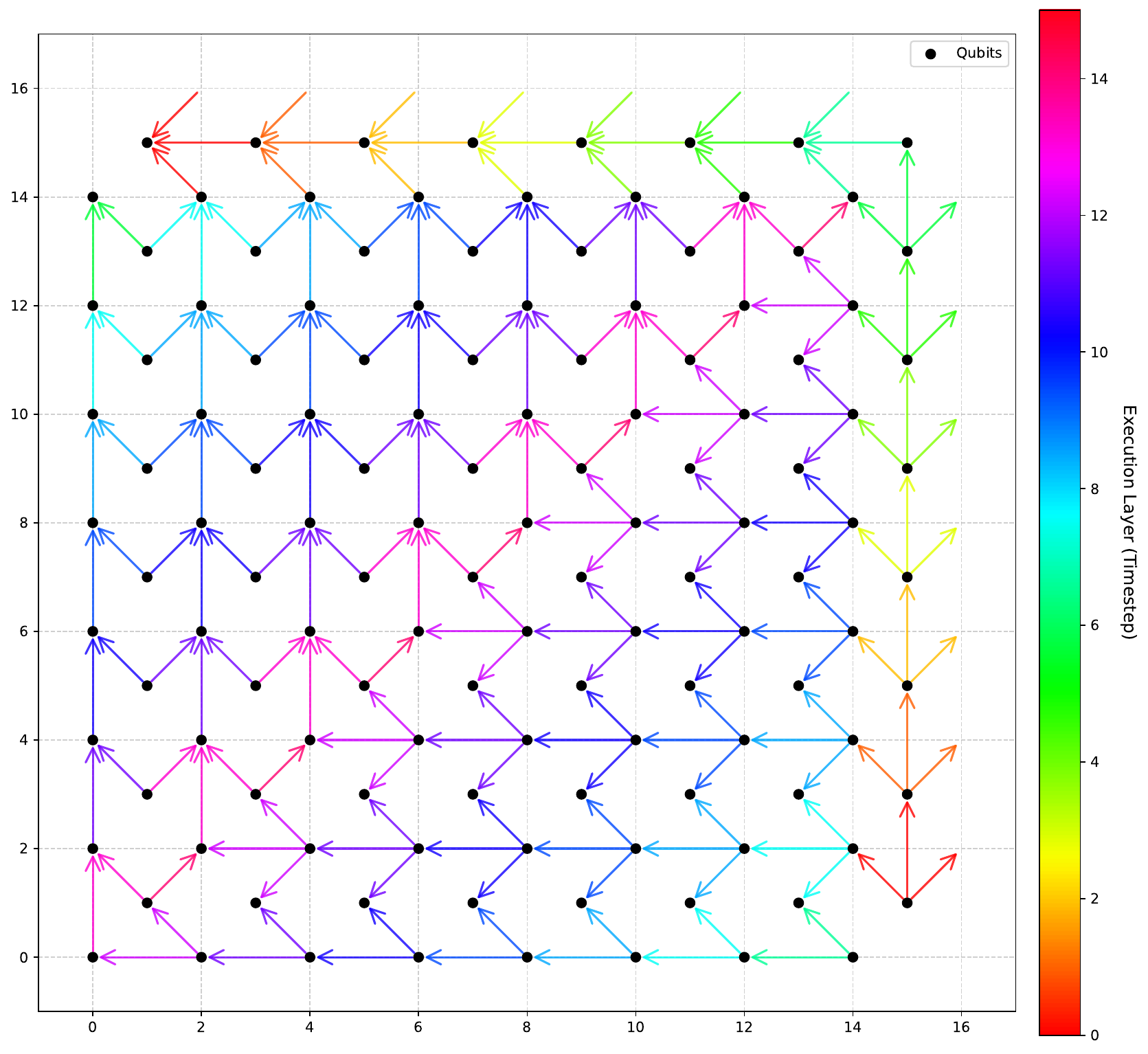}
    \caption{The output of toricCodeVisualization.py when $L=8$. The unitary $U$ is in form of $2L$ layers of commuting $CX$ gates. Each layer is in the same colour. The total gate number is $3L^2+2L-5$. }
\end{figure}

Choosing logical operators which are supported on the horizontal/vertical lines of qubits at coordinates $\{y=2L-1(X_{L_1}),x=2L-1(Z_{L_1}),x=0 (X_{L_2}),y=0(Z_{L_2})\}$, one sees that they are mapped to $(2L-1, 2L-1)$ for $X_{L_1}, Z_{L_1}$; $(0,0)$ for $X_{L_2}, Z_{L_2}$. \\

Then to encode the unknown 2 qubit state $\ket{\phi}$, applying the circuit in reverse, \[U^\dagger\ket{\phi}\bigotimes\ket{0}^{\otimes L^2-1}\bigotimes\ket{+}^{\otimes L^2-1}\] gives the encoded logical state. Since $CX^\dagger=CX$, one only needs to apply the layers of $CX$ gates in reverse. It remains to prepare the state $\ket{\phi}\bigotimes\ket{0}^{\otimes L^2-1}\bigotimes\ket{+}^{\otimes L^2-1}$ from $\ket{\phi}\bigotimes\ket{0}^{\otimes 2(L^2-1)}$, which can be done in depth 1, by applying an $H$ gate to $X$ type qubits. \\

Also note that the final Hamiltonian is of the form as described in toricCodeOperation.py:

\begin{figure}[H]
    \centering
    \includegraphics[width=10cm]{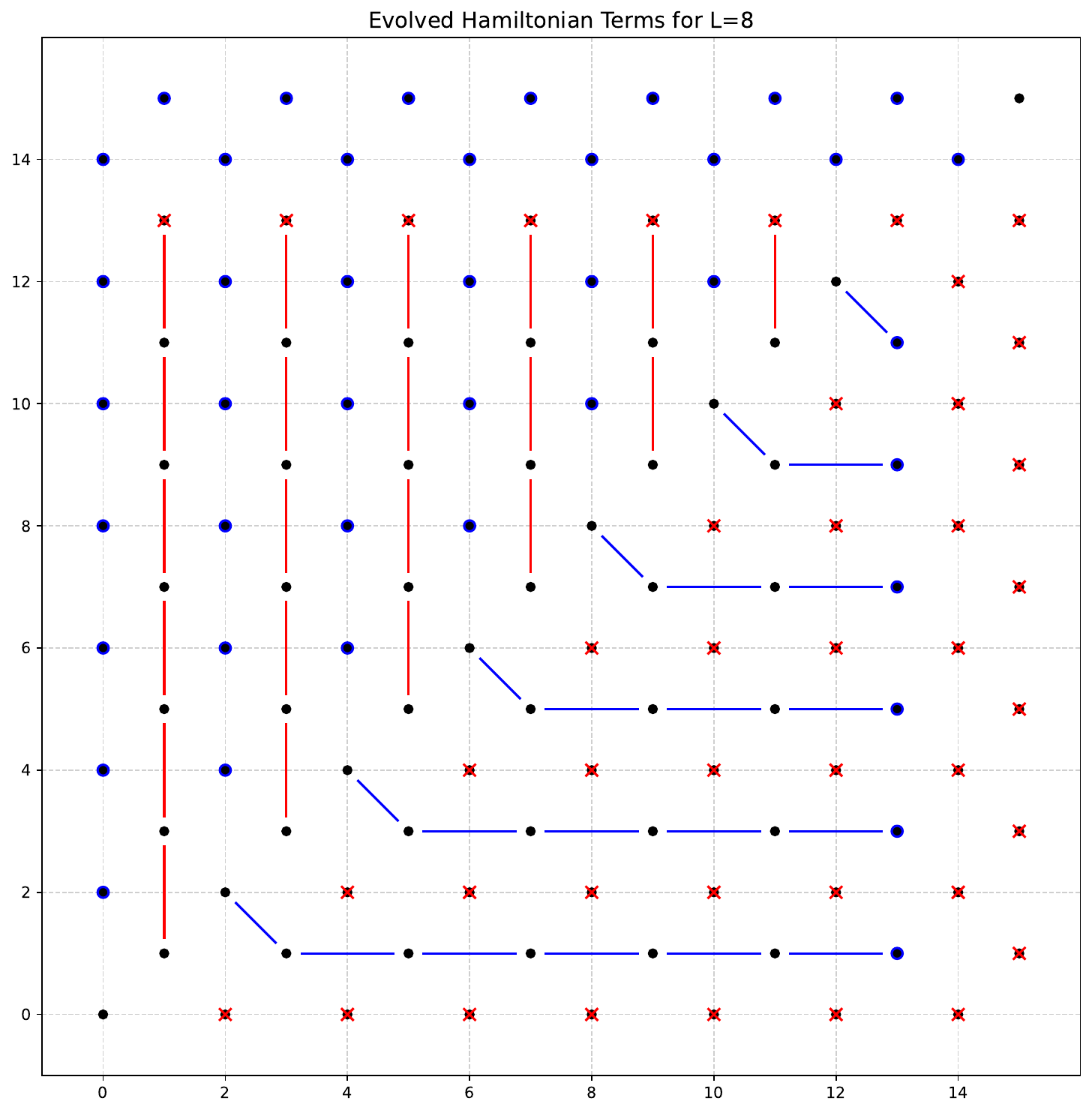}
    \caption{The output of toricCodeOperation.py when $L=8$. The trivial sites are at $(0,0),(2L-1, 2L-1)$ which are neighbours. The blue operators are $X$ interactions and the red operators are $Z$ interactions. We choose to plot $2(L^2-1)$ independent stabilizers by not plotting the 2 stabilizers that have initial support on the sites that will become trivial. }
\end{figure}

Using this choice of independent stabilizers, one observes disjoint Ising type sub-Hamiltonians with a magnetic field on one end for both $X$ and $Z$ type operators, of length $1,2,..,L-1$, and the remaining operators, which are single site, are those reduced by the algorithm directly, which means there are 3 $CX$ gates pointing to the same site or 3 $CX$ gates leaving the same site, and the four relevant sites form a rotated square. \\

\section{Conclusion and further research}
An algorithm for preparing the logical state of the toric code using local gates is presented. \\

Further improvements can be made by coming up with an algorithm that reduces the 4 logical operators simultaneously, achieving a depth of $\sim L$; it is clear that one cannot further improve beyond this asymptotic bound using the ideas above, since logical operators are required to reside at the point where they intersect with the other logical operator of different type but acting on the same site, and hence the diameter of the logical operators can only be reduced by 1 in each layer of the circuit. \\

\bibliography{references}

\end{document}